\documentclass[prl,reprint,twocolumn,footinbib,longbibliography,superscriptaddress]{revtex4-1}
\usepackage{amsmath}
\usepackage{amsfonts}
\usepackage{amssymb}
\usepackage{lmodern}
\usepackage{times}
\usepackage[dvipdfmx]{graphicx}
\usepackage[usenames,dvipsnames]{xcolor}
\usepackage{bm}
\usepackage{amsthm}
\usepackage[whole]{bxcjkjatype}

\usepackage{mathtools}
\usepackage{physics}
\usepackage{mathrsfs}
\usepackage{hyperref}

\usepackage{physics}
\usepackage{mathrsfs}

\usepackage{hyperref}
\hypersetup{
  colorlinks   = true, %Colours links instead of ugly boxes
  urlcolor     = blue, %Colour for external hyperlinks
  linkcolor    = blue, %Colour of internal links
  citecolor   = blue %Colour of citations, could be ``red''
}
\newcommand{\red}[1]{#1}

\begin{document}

\title{A geometric speed limit for acceleration by natural selection in evolutionary processes}
\date{\today}

\author{Masahiro Hoshino}
\email{hoshino-masahiro921@g.ecc.u-tokyo.ac.jp}
\affiliation{Department of Physics, The University of Tokyo, 7-3-1 Hongo, Bunkyo-ku, Tokyo 113-0033, Japan}
\author{Ryuna Nagayama}
\email{ryuna.nagayama@ubi.s.u-tokyo.ac.jp}
\affiliation{Department of Physics, The University of Tokyo, 7-3-1 Hongo, Bunkyo-ku, Tokyo 113-0033, Japan}
\affiliation{Universal Biology Institute, The University of Tokyo, 7-3-1 Hongo, Bunkyo-ku, Tokyo 113-0033, Japan}
\author{Kohei Yoshimura}
\affiliation{Department of Physics, The University of Tokyo, 7-3-1 Hongo, Bunkyo-ku, Tokyo 113-0033, Japan}
\affiliation{Universal Biology Institute, The University of Tokyo, 7-3-1 Hongo, Bunkyo-ku, Tokyo 113-0033, Japan}
\author{Jumpei F. Yamagishi}
\affiliation{Universal Biology Institute, The University of Tokyo, 7-3-1 Hongo, Bunkyo-ku, Tokyo 113-0033, Japan}
\affiliation{Graduate School of Arts and Sciences, The University of Tokyo, 3-8-1 Komaba, Meguro-ku, Tokyo 153-8902, Japan}
\author{Sosuke Ito}
\affiliation{Department of Physics, The University of Tokyo, 7-3-1 Hongo, Bunkyo-ku, Tokyo 113-0033, Japan}
\affiliation{Universal Biology Institute, The University of Tokyo, 7-3-1 Hongo, Bunkyo-ku, Tokyo 113-0033, Japan}

\begin{abstract}
    We derived a new speed limit in population dynamics, which is a fundamental limit on the evolutionary rate. By splitting the contributions of selection and mutation to the evolutionary rate, we obtained the new bound on the speed of arbitrary observables, named the selection bound, that can be tighter than the conventional Cram\'{e}r--Rao bound. Remarkably, the selection bound can be much tighter if the contribution of selection is more dominant than that of mutation. This tightness can be geometrically characterized by the correlation between the observable of interest and the growth rate. We also numerically illustrate the effectiveness of the selection bound in the transient dynamics of evolutionary processes \red{and discuss how to test our speed limit experimentally}. 
\end{abstract}

\maketitle

\textit{Introduction.---}
Biological populations fluctuate through natural selection and mutation due to various environmental influences. 
While mutation increases their diversity, natural selection increases the fraction of highly adaptive traits in the population. This competition between selection and mutation leads to evolution~\cite{kimura1983neutral,ohta1992nearly, frank2019foundations}. Recent improvements in experimental methods have enabled \red{researchers} to quantitatively observe the evolutionary dynamics of actual biological communities~\cite{arjan1999diminishing,wakamoto2005single,barrick2009genome,de2014empirical,hashimoto2016noise,baym2016spatiotemporal,de2016growth, good2017dynamics,lukavcivsinova2017toward,furusawa2018toward,van2018experimental}. For example, Ref.~\cite{baym2016spatiotemporal} visualized how selection and mutation together influence the adaptation dynamics of a bacterial population's growth. 

Though these recent experiments allow us to measure the evolutionary rate quantitatively, the classical theories for evolution were not sufficiently quantitative. For example, the principal ideas of evolution, such as natural selection in Darwinian evolution~\cite{darwin2004origin}, have not been clearly expressed quantitatively. A famous theorem on the evolutionary rate known as Fisher's fundamental theorem of natural selection~\cite{fisher1958genetical,ewens1989interpretation,frank1997price,baez2021fundamental}, which claims a relation between the increment of the mean fitness and the fitness variance, has also been misunderstood by many researchers because it is given in a quantitatively vague expression~\cite{frank1992fisher}. One exception is the Price equation~\cite{price1972,frank1992fisher, grafen2000developments,frank2012natural,baum2017selection,frank2018price,frank2020fundamental}, which provides a clear-cut relation between the observables associated with traits and their fitness. Because the Price equation is a purely mathematical relation based on identity, we need to consider specific population dynamics~\cite{malthus1872,leibler2010individual,crow2017introduction,basener2018fundamental} to identify its physical implication for the evolutionary rate. 

Recently, quantitative theoretical approaches have been developed by analogy with another developing field of stochastic thermodynamics~\cite{sekimoto2010stochastic,seifert2012stochastic}. \red{In population dynamics models} such as the Lotka--Volterra model~\cite{strogatz2018nonlinear} and the lineage trees~\cite{wakamoto2012optimal,lambert2015quantifying,hoffmann2016nonparametric,nozoe2017inferring}, several quantitative inequalities or trade-off relations for the evolutionary processes have been investigated~\cite{andrae2010entropy,qian2014fitness, kobayashi2015fluctuation,sughiyama2015pathwise, sughiyama2017steady,kobayashi2017stochastic, garcia2019linking,  genthon2020fluctuation,genthon2021universal, yoshimura2021thermodynamic,kolchinsky2021thermodynamic} by analogy with thermodynamic laws such as the second law of thermodynamics~\cite{esposito2010three,van2010three} and thermodynamic uncertainty relations~\cite{barato2015thermodynamic}. As a notable result, \red{the evolutionary rate has been discussed quantitatively} in Ref.~\cite{zhang2020information} by applying the information-geometric speed \red{limits~\cite{crooks2007measuring, ito2018stochastic}. The speed limits have} been discussed as a classical counterpart of the quantum speed limits~\cite{mandelstam1991uncertainty,anandan1990geometry,margolus1998maximum, taddei2013quantum,pires2016generalized,shanahan2018quantum, okuyama2018quantum} in the context of a connection between information geometry~\cite{amari2016information} and stochastic thermodynamics. \red{As a constraint on the speed of dynamics, the speed limits offer a basis to} discuss the evolutionary rate quantitatively. \red{These speed limits have} also been generalized \red{to} the speed of observable~\cite{ito2020stochastic,nicholson2020time, yoshimura2021information,ito2022information, ashida2021experimental} based on the Cram\'{e}r--Rao bound~\cite{amari2016information, rao1992information}, well known in information geometry. 
\red{This} information-geometric approach would be promising as a quantitative theory for the evolutionary rate \red{because of a deep connection between the Cram\'{e}r--Rao bound and the Price equation~\cite{frank2018price, frank2020fundamental} and because this approach may be compatible with the existing} information-theoretic and stochastic methods for evolutionary dynamics~\cite{sato2003relation,michel2005general,wolf2005diversity, kussell2005phenotypic, rivoire2011value,frank2012naturalinfo,frank2013natural,kaneko2015universal, xue2018benefits, nakashima2022acceleration}. Indeed, several applications and generalizations of the speed limits have been recently studied to understand the speed in population dynamics quantitatively~\cite{adachi2022universal,garcia2022diversity}.

\red{However, those previous studies~\cite{zhang2020information, adachi2022universal, garcia2022diversity} did not focus on the competing situations of natural selection and mutation, even though selection and mutation together shape evolution.} % \gray{Ref.~\cite{adachi2022universal} mainly focused on universal aspects of speed limits for various population dynamics. Though Ref.~\cite{zhang2020information} discussed the effect of natural selection and Ref.~\cite{garcia2022diversity} showed the effect of the difference between mutation-driven and mutation-less dynamics on the evolutionary rate, they did not discuss details on the competing situations of natural selection and mutation.} 
\red{Here, we pose} the following unresolved issue: {\it how and when does the evolutionary rate change in the competing situations of natural selection and mutation?} This question would be crucial for a quantitative understanding of evolutionary processes where the competition between selection and mutation \red{can} enhance the evolutionary rate, as observed in Ref.~\cite{baym2016spatiotemporal}. 

To resolve such an issue, we \red{theoretically evaluated} the evolutionary rate \red{by decomposing it} into the contributions of natural selection and mutation in population dynamics\red{, thereby deriving a new speed limit}. This speed limit is tighter than the conventional Cram\'{e}r--Rao bound when natural selection is dominant \red{compared to mutation (e.g., in transient dynamics of evolution), as analytically proven and numerically illustrated. It describes} how natural selection accelerates evolution. % \red{compared to the case where only mutations occur}. 
% We show that the new speed limit is effective when the contribution of natural selection is significant compared to that of mutation in transient dynamics of evolution. We also analytically obtain the geometric condition of this effectiveness and illustrate it by numerical calculations.

\begin{figure}[t]
    \centering
    \includegraphics[scale=0.78]{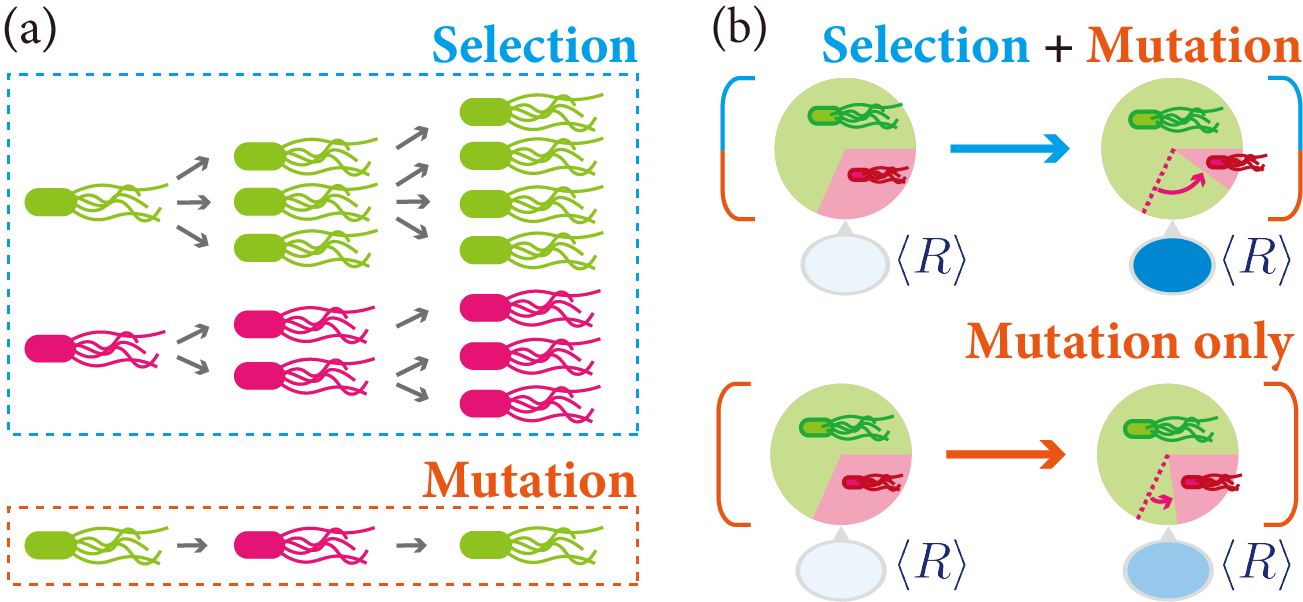}
    \caption{(a) Schematic illustration of the model. Population size changes due to selection on growth and mutation. (b) Schematic illustration of the selection bound. Compared to the case where the population changes only due to mutation, the change of $\langle R \rangle$ can be accelerated when both selection and mutation affect the population dynamics. }
    \label{fig:conceptart}
\end{figure}

\textit{Setup.---}
To discuss speed limits for observables in population dynamics, we consider a model consisting of selection and mutation between multiple traits~\cite{crow2017introduction} (see also Fig.~\ref{fig:conceptart}(a)). Suppose a population consists of subpopulations with $n$ different traits, and $N_{i}(t)$ and $\lambda_{i}$ denote the number and growth rate of individuals in the subpopulation with the $i$-th trait at time $t$, respectively. The traits may be phenotypic, genotypic, or epigenetic properties. We denote the vector of growth rates simply as $\lambda\,{=}\,(\lambda_i)$. Mutation is assumed to be a Markovian process with transition rate matrix $\mathsf{W}\,{=}\,(W_{ij})$, where the $(i,j)$-element $W_{ij}$ indicates the transition rate from trait $j$ to $i$ if $i\,{\neq}\,j$, and the elements satisfy $\sum_{i=1}^{n}W_{ij}{=}\,0$ and $W_{ij}\,{\geq}\, 0\;(i\,{\neq}\,j)$. We assume that $\lambda$ and $\mathsf{W}$ are time-independent \red{because we consider a stationary environment.} 
Then $N_{i}(t)$ follows the following differential equation
\begin{align}\label{eq.N_dynamics}
    \frac{d}{dt}N_i(t)=\lambda_{i}N_i(t)+\sum_{j=1}^{n}W_{ij}N_{j}(t).
\end{align}
The first term on the right-hand side represents the change in the population due to selection and the second term represents the change in the population due to mutation. 
With $N_{\rm tot}(t)\,{\coloneqq}\, \sum_{i=1}^{n}N_i(t)$, the proportion $p_i(t)\,{\coloneqq}\, N_i(t)/N_{\rm tot}(t)$ of each subpopulation satisfies the definition of the probability distribution, i.e., the non-negativity $p_i(t)\,{\geq}\,0$ and the normalization $\sum_{i=1}^{n}p_i(t)\,{=}\,1$.　
From Eq.~\eqref{eq.N_dynamics}, this ``probability distribution'' follows a nonlinear master equation,
\begin{align}\label{eq.p_dynamics}
    \frac{d}{dt}p_i(t)=\Delta \lambda_{i} p_i(t)+\sum_{j=1}^{n}W_{ij}p_{j}(t),
\end{align}
where the ensemble average of an observable $A\,{=}\,\{A_i\}_{i=1}^{n}$ with respect to $p_i(t)$ is defined as $\langle A\rangle\,{\coloneqq}\, \sum_{i=1}^{n}p_i(t)A_i$ and $\Delta A_i\,{=}\, A_i - \langle A\rangle$ denotes the deviation of $A_i$.

\textit{Speed limit and information geometry.---}
We here briefly explain the conventional \red{information-geometric} speed limit for a time-independent observable $R\,{=}\,\{R_i\}_{i=1}^{n}$. The speed of observable $R$ is defined as 
\begin{align}\label{eq:vrdef}
    v_R\coloneqq \frac{1}{\sqrt{\mathrm{Var}[R]}}\frac{d\langle R\rangle}{dt}, 
\end{align}
where $\mathrm{Var}[R]\,{\coloneqq}\, \langle (\Delta R)^2 \rangle$ is the variance of $R$. In population dynamics, $v_R$ quantifies the evolutionary rate with respect to observable $R$. For example, \red{the evolutionary rate with respect to the growth rate, $v_{\lambda}$, is} given by the time derivative of the averaged growth rate $d\langle \lambda \rangle/dt$ normalized by its standard deviation $\sqrt{\mathrm{Var}[\lambda]}$. 
The speed limit for observable, known as the Cram\'{e}r--Rao bound, is a universal constraint on this speed $v_R$ for any observable  $R$~\cite{ito2020stochastic,nicholson2020time}: 
\begin{align}\label{eq.speedlimit.normal}
    -v_{\rm info}\leq v_R\leq v_{\rm info},
\end{align}
which holds for arbitrary dynamics of \red{a probability distribution~\footnote{We remark that the Cram\'{e}r--Rao bound holds \red{even} for time-dependent observables by changing the definition of $v_R$ as $v_R\,{=}\,(d_t\langle R\rangle-\langle d_tR\rangle)/\sqrt{\mathrm{Var}[R]}$.}.} 
Here $v_{\rm info}$ \red{is defined as} the square root of the Fisher information~\cite{cover1999elements,frank2009natural,ito2020stochastic}:
\begin{align}\label{def.fisherinformation}
    v_{\rm info} \coloneqq \sqrt{\sum_{i=1}^{n}p_i\qty(\frac{d\ln p_i}{dt})^2}.
\end{align}
In information geometry, \red{we can interpret it as} the speed of a probability distribution moving on a manifold of distributions.

\textit{Fitness and Price equation.---}
The square root of the Fisher information $v_{\rm info}$ not only indicates the speed of the probability distribution but also characterizes the population dynamics \red{because it is identified with} the variance of fitness~\cite{frank2018price,baez2021fundamental}. We here introduce the fitness $f_i$ of trait $i$ as the effective growth rate of $N_i$:
\begin{align}\label{def.fitness}
    f_i\coloneqq \frac{d}{dt}\ln N_i(t). 
\end{align}
The ensemble average of the fitness is equal to the effective growth rate of the total population: $\langle f\rangle\,{=}\,d \ln N_{\rm tot}/dt$. Together with Eq.~\eqref{def.fisherinformation} and $\ln p_i(t) \,{=}\, \ln N_i(t)-\ln N_{\rm tot}(t)$, these relations lead to the equality
\begin{align}
   v_{\rm info} = \sqrt{ \sum_{i=1}^{n}p_i\qty(\Delta f_i)^2 }= \sqrt{\mathrm{Var}[f]}.
   \label{variancefitness}
\end{align}
\red{That is, $v_{\rm info}$ also} quantifies the diversity of each trait's fitness.
Accordingly, Eq.~\eqref{eq.speedlimit.normal} implies that the variance of fitness limits the % change 
speed of an arbitrary observable.

On the other hand, we can discuss \red{the role of} $v_R$ in population dynamics based on the Price equation~\cite{price1972,frank1992fisher, grafen2000developments,baum2017selection,frank2018price,frank2020fundamental,frank2012natural}. A special case of the Price equation for a time-independent observable provides a connection with the time derivative of the stochastic entropy in the system, $\dot{\sigma}_i(t)\,{\coloneqq}\, -d\ln p_i(t)/dt$, as
\begin{align}\label{eq.pricealign}
    \frac{d\langle R\rangle}{dt} = \mathrm{Cov}[R,-\dot{\sigma}],
\end{align}
where the covariance of two observables is defined as $\mathrm{Cov}[A,B]\,{\coloneqq}\,\langle \Delta A \Delta B\rangle$.
This equation indicates that the evolutionary rate is governed by the stochastic entropy change rate in the system. 
We remark that $\dot{\sigma}$ is directly connected to the fitness as $-\dot{\sigma}_i\,{=}\, \Delta f_i$, so that its ensemble average and variance satisfy $\langle\dot{\sigma}\rangle\,{=}\,0$ and $\sqrt{\mathrm{Var}[\dot{\sigma}}]\,{=}\,\sqrt{\mathrm{Var}[f]}\,{=}\,v_{\rm info}$. From Eq.~\eqref{eq.pricealign}, $v_R$ is rewritten as
\begin{align}\label{def.v_R}
v_R= \frac{\mathrm{Cov}[R,-\dot{\sigma}]}{\sqrt{\mathrm{Var}[R]}} = \frac{\mathrm{Cov}[R,f]}{\sqrt{\mathrm{Var}[R]}},
\end{align}
which implies that the speed of an observable can be interpreted in terms of the covariance between the observable and the fitness. From Eqs.~(\ref{variancefitness}) and (\ref{def.v_R}), the Cram\'{e}r--Rao bound~\eqref{eq.speedlimit.normal} can be derived by applying the Cauchy--Schwarz inequality $- \sqrt{\mathrm{Var}[R]}\sqrt{\mathrm{Var}[f]} \,{\leq}\, \mathrm{Cov}[R,f]\,{\leq}\, \sqrt{\mathrm{Var}[R]}\sqrt{\mathrm{Var}[f]}$.

\textit{Main result: Selection bound.---} We explain the main result which is a new speed limit based on the contribution of selection in evolutionary dynamics. The key idea for the main result is the decomposition of the stochastic entropy change rate in the system. In the population dynamics model~\eqref{eq.p_dynamics}, $\dot{\sigma}$ can be decomposed into two parts as
\begin{align}\label{eq.entropy.div}
    \dot\sigma = {\dot\sigma}^{\lambda} + {\dot\sigma}^{\mathsf{W}},
\end{align}
where ${\dot\sigma}^{\lambda}_i \,{\coloneqq}\, - \Delta \lambda_i$ and ${\dot\sigma}^{\mathsf{W}}_i\, {\coloneqq}\,-\sum_{j=1}^{n}W_{ij}p_j/p_i$ are the stochastic entropy change rate in the system due to only selection and mutation, respectively. We remark that ${\dot\sigma}^{\mathsf{W}}$ is rewritten as ${\dot\sigma}^{\mathsf{W}}_i \,{=}\,- \Delta (f_i- \lambda_i)$. 
These quantities also satisfy $\langle\dot{\sigma}^\lambda\rangle\,{=}\,\langle\dot{\sigma}^{\mathsf{W}}\rangle\,{=}\,0$, as $\dot{\sigma}$ does. Considering this decomposition, we introduce the following quantities,
\begin{equation}
    \begin{split}
        v_R^\lambda \coloneqq \dfrac{\mathrm{Cov}[R,-\dot{\sigma}^\lambda]}{\sqrt{\mathrm{Var}[R]}},\quad & 
        v_{\rm info}^\lambda \coloneqq \sqrt{\mathrm{Var}[\dot{\sigma}^\lambda]},\\
        v_R^{\mathsf{W}} \coloneqq \dfrac{\mathrm{Cov}[R,-\dot{\sigma}^{\mathsf{W}}]}{\sqrt{\mathrm{Var}[R]}},\quad&
        v_{\rm info}^{\mathsf{W}} \coloneqq \sqrt{\mathrm{Var}[\dot{\sigma}^{\mathsf{W}}]},
    \end{split}
\end{equation}
where the upper two can be interpreted as $v_R$ and $v_{\rm info}$ \red{without} the contribution of mutation, while the lower ones are interpreted as those without selection. In other words, the former are speeds stemming solely from selection, while the latter mutation only. 
These quantities are given by the variance and covariance of the measurable observables $R$, $\lambda$, and $f- \lambda$; 
$v_R^\lambda \,{=}\,\mathrm{Cov}[R,\lambda]/\sqrt{\mathrm{Var}[R]}$, 
$v_{\rm info}^\lambda\,{=}\,\sqrt{\mathrm{Var}[\lambda]}$, $v_R^{\mathsf{W}} \,{=}\,\mathrm{Cov}[R,f-\lambda]/\sqrt{\mathrm{Var}[R]}$, 
and $v_{\rm info}^{\mathsf{W}} \,{=}\,\sqrt{\mathrm{Var}[f-\lambda]}$. 
By decomposing $v_R$ into the contributions of selection and mutation, we obtain a new speed limit:
\begin{align}\label{eq.naturalselectionbound}
v_{R}^{\lambda}-v_{\rm info}^{\mathsf{W}} 
     \leq v_R 
     \leq v_{R}^{\lambda}+v_{\rm info}^{\mathsf{W}} .
\end{align}
We call this new speed limit \textit{the selection bound} because the speed of observable $v_R$ is accelerated by the effect of selection  $v_R^{\lambda}$, compared to the case where no selection occurs ($-v_{\rm info}^{\mathsf{W}} \,{\leq}\, v_R \,{\leq}\, v_{\rm info}^{\mathsf{W}}$ for $\lambda \,{=}\,0$). 
Therefore, this bound quantifies the acceleration of the evolutionary rate by natural selection compared to mutational dynamics in the absence of the selection (see also Fig.~\ref{fig:conceptart}(b)).

The equation~\eqref{eq.naturalselectionbound} is derived essentially from the Cauchy--Schwarz inequality as well as the Cram\'{e}r--Rao bound. To simplify its derivation, we define an inner product and the associated norm for observables as $\langle A,B\rangle\,{\coloneqq}\,\sum_{i=1}^{n}p_iA_iB_i$ and $\Vert A\Vert\,{:=}\,\sqrt{\langle A,A\rangle}$, respectively.  We can rewrite $v_R$ as
\begin{align}\label{eq.vR.div}
    v_R 
    &= \dfrac{\langle\Delta R, \Delta \lambda+ \Delta (f- \lambda)\rangle}{\Vert \Delta R\Vert} = v_R^\lambda + v_R^{\mathsf{W}}.
\end{align}
Applying the Cauchy--Schwarz inequality $- \Vert \Delta R\Vert \Vert \Delta (f- \lambda)\Vert \,{\leq}\, \langle\Delta R, \Delta (f- \lambda)\rangle\,{\leq}\, \Vert \Delta R\Vert \Vert \Delta (f- \lambda)\Vert$ to Eq.~\eqref{eq.vR.div}, we can obtain the selection bound. 

\begin{figure}[t]
    \centering
    \includegraphics[width=\linewidth]{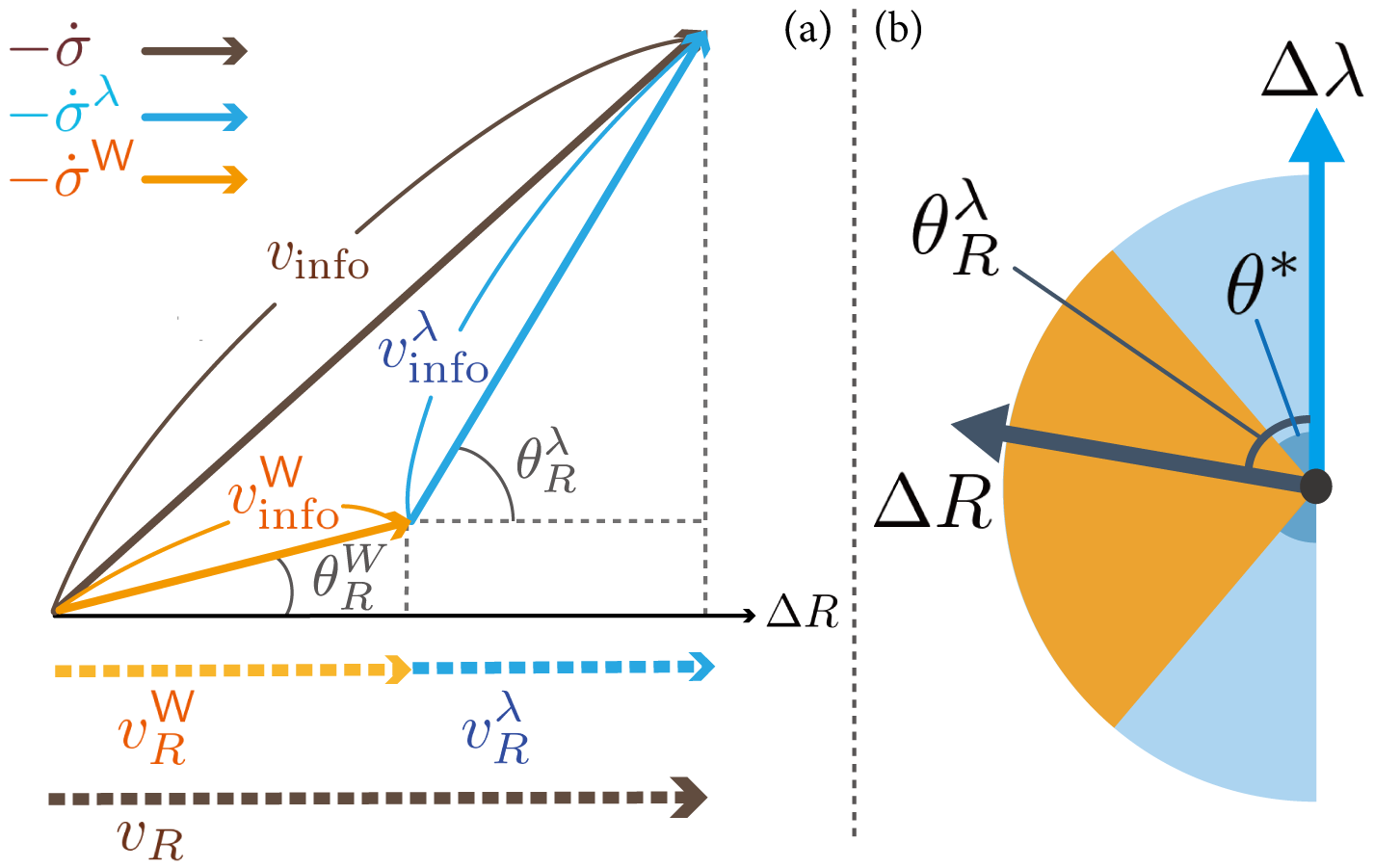}
    \caption{(a) Relations between the quantities presented in this Letter in the inner product space. $-\dot\sigma$ can be written by the sum of the contributions of selection and mutation, $-{\dot\sigma}^{\lambda}$ and $-{\dot\sigma}^{\mathsf{W}}$. The norms of $-\dot\sigma, -{\dot\sigma}^{\lambda},-{\dot\sigma}^{\mathsf{W}}$ are $v_{\rm info},v_{\rm info}^{\lambda},v_{\rm info}^{\mathsf{W}}$, respectively. The speed $v_R$ can be expressed by the projection of $-\dot\sigma$ in the direction of $\Delta R$, while the projections of $-{\dot\sigma}^{\lambda}$ and $-{\dot\sigma}^{\mathsf{W}}$ are $v_R^{\lambda}$ and $v_R^{\mathsf{W}}$, respectively. Thus, $v_R=v_R^{\lambda}+v_R^{\mathsf{W}}$ holds. (b) The range of $\theta_R^{\lambda}$ that determines whether the selection bound or the Cram\'{e}r--Rao bound evaluates $v_R$ tightly or loosely. If $\theta^\ast\,{\leq}\, \theta_R^{\lambda}\,{\leq}\, \pi$, the selection bound gives a tighter upper bound, and if $0\,{\leq}\, \theta_R^{\lambda}\,{\leq}\, \pi-\theta^\ast$, the selection bound gives a tighter lower bound. Therefore, both upper and lower bounds are tight when $\theta_R^{\lambda}$ is located in the orange area $[\theta^\ast,\pi-\theta^\ast]$.}
    \label{fig:geometric}
\end{figure}

This inner product also provides a useful geometric interpretation to discuss the effectiveness of the selection bound (see also Fig.~\ref{fig:geometric}(a)). In \red{the} geometric interpretation, $v_\mathrm{info}\,{=}\,\Vert \dot{\sigma}\Vert$, $v_\mathrm{info}^\lambda\,{=}\,\Vert \dot{\sigma}^\lambda\Vert$ and $v_\mathrm{info}^\mathsf{W}\,{=}\,\Vert \dot{\sigma}^\mathsf{W}\Vert$ are the norms of $-\dot{\sigma}$, $-\dot{\sigma}^\lambda$ and $-\dot{\sigma}^\mathsf{W}$, \red{respectively, and thus the triangle inequality $\abs{v_{\rm info}^{\lambda}-v_{\rm info}^{\mathsf{W}}}\,{\leq}\, v_{\rm info}\,{\leq}\, v_{\rm info}^{\lambda}+v_{\rm info}^{\mathsf{W}}$ holds. In addition,} $v_R\,{=}\, \langle\Delta R,-\dot\sigma\rangle/ \Vert\Delta R\Vert$, $v_R^\lambda\,{=}\, \langle\Delta R,-\dot\sigma^\lambda\rangle/ \Vert\Delta R\Vert$, and $v_R^\mathsf{W}\,{=}\, \langle\Delta R,-\dot\sigma^\mathsf{W}\rangle/ \Vert\Delta R\Vert$ are the norms of the projections of $-\dot{\sigma}$, $-\dot{\sigma}^\lambda$ and $-\dot{\sigma}^\mathsf{W}$ onto $\Delta R$, respectively. 
\red{Then,} the angle between $\Delta R$ and $-\dot{\sigma}^\lambda\,{=}\,\Delta \lambda$ \red{can be defined} as 
\begin{align}
    \theta_R^{\lambda}
    \coloneqq\arccos(
    \frac{\langle\Delta R, \Delta \lambda\rangle}
    {\Vert\Delta R\Vert
    \Vert\Delta \lambda\Vert})
    =\arccos(\frac{v_R^{\lambda}}{v_{\rm info}^{\lambda}}),
\end{align}
or equivalently $v_R^\lambda\,{=}\,v_{\rm info}^\lambda\cos\theta_R^\lambda$\red{. It} quantifies the strength of the correlation between $R$ and $\lambda$~\red{\footnote{\red{We remark that a similar concept of the angle can be seen in the Price equation~\cite{frank2012natural}.}}}. % \gray{This interpretation clarifies the condition where the selection bound is effective.} 
If there is a positive correlation between an observable $R$ and the growth rate $\lambda$ (i.e., $\cos\theta_R^{\lambda}\,{\geq}\,0$), $v_R^{\lambda}$ is positive and both upper and lower bounds in the selection bound shift to the positive direction (i.e., $\pm v_{\rm info}^{\mathsf{W}} + v_R^{\lambda}$), compared to the case without selection (i.e., $\pm v_{\rm info}^{\mathsf{W}}$). Therefore, positive correlations between observables and the growth rate can lead to faster evolution. From a biological viewpoint, it indicates that if the value of the observable $R$ tends to be larger in fast-growing traits, its evolutionary rate can be accelerated, and vice versa. 
\red{Noting the relation $v_R^\lambda\,{=}\,v_{\rm info}^\lambda\cos\theta_R^\lambda$, not only stronger correlations between observables and growth rate (i.e., larger $\cos\theta_R^\lambda$) but also greater contributions of selection (i.e., larger $v_{\rm info}^\lambda$, as discussed above) together allow a faster evolutionary rate.} % At the same time, we can find that $v_{\rm info}^\lambda$ also affects the acceleration from the relation $v_R^\lambda\,{=}\,v_{\rm info}^\lambda\cos\theta_R^\lambda$. Noting that $v_{\rm info}^\lambda$ quantifies the contribution of selection as discussed above, stronger correlations between observables and growth rate and more significant contribution of selection can lead to a faster evolutionary rate.

Finally, let us compare our bound~\eqref{eq.naturalselectionbound} with the conventional Cram\'{e}r--Rao bound~\eqref{eq.speedlimit.normal} to see how it quantifies the competition between selection and mutation. 
To this end, we define another angle $\theta^\ast$ as
\begin{align}\label{def.theta_ast}
    \theta^\ast \coloneqq \arccos{\qty(\dfrac{v_{\rm info} - v_{\rm info}^{\mathsf{W}}}{v_{\rm info}^{\lambda}})},
\end{align}
which is well-defined because the argument of the arccosine is always in $[-1,1]$ from the triangle inequality. Using $\theta_R^\lambda$ and $\theta^\ast$, the following case separation gives the condition in which case the Cram\'{e}r--Rao bound or the selection bound gives better evaluation: 
\begin{align}\label{eq.bound.eval}
    \begin{cases}
        0\leq \theta_R^{\lambda}\leq \theta^\ast &\implies v_R\leq v_{\rm info}\leq  v_R^{\lambda}+v_{\rm info}^{\mathsf{W}}\\
        \theta^\ast\leq \theta_R^{\lambda}\leq \pi &\implies v_R\leq v_R^{\lambda}+v_{\rm info}^{\mathsf{W}} \leq v_{\rm info}  \\
        0\leq \theta_R^{\lambda}\leq \pi-\theta^\ast &\implies -v_{\rm info} \leq v_R^{\lambda}-v_{\rm info}^{\mathsf{W}}\leq v_R\\
        \pi -\theta^\ast\leq \theta_R^{\lambda}\leq \pi &\implies v_R^{\lambda}-v_{\rm info}^{\mathsf{W}}\leq -v_{\rm info}\leq v_R
    \end{cases}.
\end{align}
This implies that if $\theta^\ast$ is smaller than $\pi/2$ and \red{$\theta_R^{\lambda}$} is in the range $[\theta^\ast,\pi-\theta^\ast]$, 
then the selection bound will bound $v_R$ more tightly than the Cram\'{e}r--Rao bound, both lower and above (see Fig.~\ref{fig:geometric}(b)). 
Since this range $[\theta^\ast,\pi-\theta^\ast]$ does not depend on the choice of specific observables, we can discuss the tightness of the selection bound quantitatively only by the angles. 
\red{Because $\theta^\ast$ is given as the arccosine of the ratio between $v_\mathrm{info}-v_\mathrm{info}^\mathsf{W}$ and $v_\mathrm{info}^\lambda$ and $v_{\rm info}$ is given as the sum of $v_{\rm info}^\lambda$, $v_{\rm info}^{\mathsf{W}}$ and a correlation term between them (cf. \textit{cosine theorem}), $\theta^\ast$ becomes smaller when the contribution of selection $v_{\rm info}^\lambda$ gets larger, since the ratio gets closer to one. }
\red{That is, if the contribution of selection is larger than mutation ($v_\mathrm{info}^\lambda\gg v_\mathrm{info}^\mathsf{W}$), 
the range becomes wider so that the selection bound can give a better bound on $v_R$ for a wider variety of observables $R$.} 
Such a tendency is indeed observed in the numerical calculations below.

\begin{figure}[t]
    \centering
    \includegraphics[width=\columnwidth]{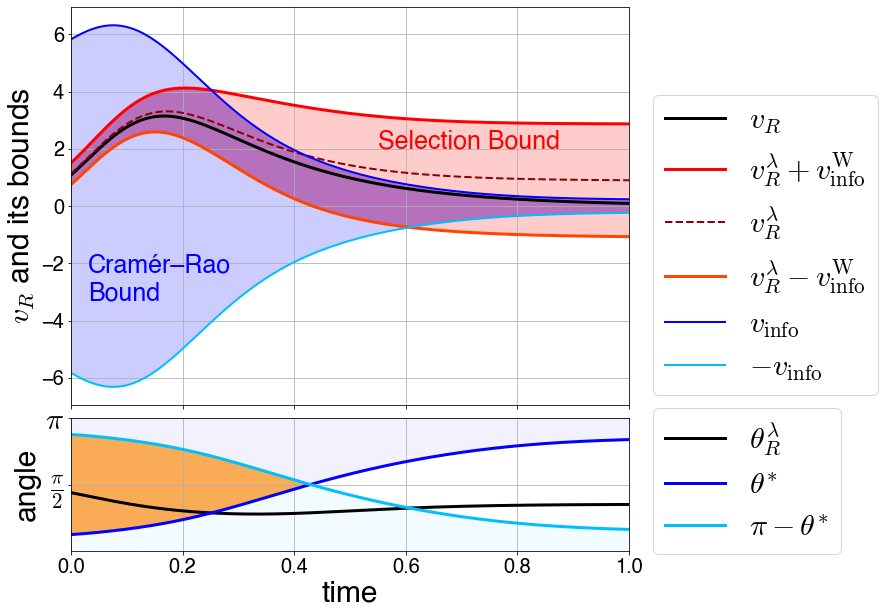}
    \caption{Numerical calculation for the conventional speed limit~\eqref{eq.speedlimit.normal} by the Cram\'{e}r--Rao bound and the new speed limit~\eqref{eq.naturalselectionbound} by the selection bound. The horizontal axis is time, and $v_R,\;v_R^{\mathsf{W}}+v_{\rm info}^{\lambda},\;v_R^{\mathsf{W}}-v_{\rm info}^{\lambda},\;v_{\rm info}$, and $-v_{\rm info}$ are plotted in the upper row.  The lower row shows the angles, $\theta_{R}^{\lambda},\;\theta^\ast,$ and $\pi-\theta^\ast$. }
    \label{fig.NS.graph}
\end{figure}

\textit{Example.---}
We illustrate our results by numerical calculations (Fig.~\ref{fig.NS.graph}). 
To consider a situation where the contribution of selection is dominant, \red{we have the parameters in~\eqref{eq.p_dynamics}, 
$\{\lambda_i\}$ and $\{W_{ij}\}_{i\neq j}$, uniformly sampled from $[-\lambda_{\rm max},\lambda_{\rm max}]$ and $[0,\mathsf{W}_{\rm max}]$, and set $\lambda_{\rm max}/\mathsf{W}_{\rm max}=100$. }
Given that the present results hold for arbitrary time-independent observables, \red{we also uniformly sample $\{R_i\}$ within the range} $[-10,10]$ rather than taking a specific observable. 
In Fig.~\ref{fig.NS.graph}, at an early stage of the evolutionary dynamics, or far from the steady state, the selection bound better restricts $v_R$. 
\red{From the ecological perspective, this behavior seems reasonable: selection dominantly contributes to the evolutionary processes far from steady states because beneficial mutation gets less likely as evolution progresses.} % It is reasonable because selection dominantly contributes to the evolutionary processes far from steady states. \textcolor{red}{We can also interpret this result from the ecological perspective as mutation increasing the ensemble average of the fitness is less likely to occur as evolution progresses.}
As proven above, the selection bound is tighter than the Cram\'{e}r--Rao bound when $\theta_R^\lambda$ is in the range $[\theta^\ast,\pi-\theta^\ast]$ (colored in orange in Fig.~\ref{fig.NS.graph}). 
This range is wider when the selection is dominant. 
\red{A more precise evaluation of the speed limits is discussed in Supplemental Material (SM)~\cite{SuppleMaterial}.}

\textit{Complementary result: Mutation bound.---}
The discussion so far has focused on how the evolutionary rate is accelerated by selection on growth. On the other hand, we can discuss acceleration by mutation by inverting the roles of $\mathsf{W}$ and $\lambda$ in the selection bound. Concretely, we can derive the bound
\begin{align}
    v_R^{\mathsf{W}}-v_{\rm info}^{\lambda} \leq v_R \leq v_R^{\mathsf{W}}+v_{\rm info}^{\lambda}.
\end{align}
We call this bound \textit{the mutation bound} because it extracts the effect of mutation $v_R^{\mathsf{W}}$. The same analysis can be performed for the mutation bound as for the selection bound. With both the selection bound and the mutation bound, we can better capture the characteristics of evolutionary processes, especially in the competing situation of natural selection and mutation (see SM~\cite{SuppleMaterial} for details).

\textit{\red{Experimental accessibility.---}}\red{Our speed limit is quantitatively testable by actual experiments using single-cell lineage tree data. Recent advances in experimental techniques enable us to measure when and into what each cell mutates or divides. For example, by analyzing time-lapse images of growing bacteria~\cite{nozoe2017inferring}, we can measure transitions in phenotypic traits (e.g., cell sizes, shapes, and intracellular concentration of a particular protein) and proliferation dynamics at the same time. The number of individuals with $i$-th trait at time $t$ that have experienced $K$ divisions since time $t'$, denoted as $N_i(K,t;t')$, can be obtained in such an experiment. This quantity $N_i(K,t;t')$ enables us to compute the instantaneous values of all the quantities in our speed limit, $p_i(t)$, ${\dot\sigma}_i^{\lambda}$, ${\dot\sigma}_i^{\mathsf{W}}$, ${\dot\sigma}_i^{\lambda}$, $v_{R}^{\lambda}, v_{R}^{\mathsf{W}}, v_{\rm info}^{\lambda}$, and $v_{\rm info}^{\mathsf{W}}$ (see SM~\cite{SuppleMaterial}). Therefore, we can experimentally check the tightness of our speed limit for arbitrary observable $R$, which quantifies the contribution of observable $R$ to the evolutionary rate in the selection process.}

\textit{Conclusion.---}
We derived a novel speed limit, the selection bound, that considers the contributions of the selection and mutation separately when the competition between selection and mutation exists. The core of this result is the decomposition of the stochastic entropy change rate into the selection and mutation part. It allows us to understand the limitations of the speeds of observables more precisely than the conventional speed limit from the Cram\'{e}r--Rao bound. Although the limits from the selection bound depend on the observables we consider, the ``tendency'' for the selection bound to give a better bound than the Cram\'{e}r--Rao bound only depends on the selection strength. The selection bound should be effective in selection-dominant situations such as environmental shift conditions. 

\red{The decomposition of the stochastic entropy change rate in this Letter may be applicable to other nonlinear dynamics. For example, the generalized Lindblad equation for post-selection in quantum dynamics~\cite{wiseman1994quantum,cresser2006measurement,zhou2022generalized} has a similar nonlinear term originated by the normalization of a probability distribution. 
Our decomposition into a nonlinear contribution (i.e., selection) and a linear contribution (i.e., mutation) may be generalized for such an equation to derive a specialized speed limit that characterizes the property of post-selection.}

\begin{acknowledgments}
M.~H., J.~F.~Y., and S. I thank Shion Orii for valuable discussions. R.~N. and S.~I. thank Chikara Furusawa and Yusuke Himekoka for helpful comments. S.~I. and J.~F.~Y. also thank Kouki Yamada for valuable discussions. 
S. I. is supported by JSPS KAKENHI Grants No.\ 19H05796, No.\ 21H01560, and No.\ 22H01141, JST Presto Grant No.\ JPMJPR18M2, and UTEC-UTokyo FSI Research Grant Program. 
K.~Y.\ and J.~F.~Y.\ are supported by Grant-in-Aid for JSPS Fellows (Grant No.~22J21619 and No.~21J22920, respectively).

M.~H. and R.~N. contributed equally to this work.
\end{acknowledgments}

\clearpage
\onecolumngrid
\appendix

\begin{center}
    {\large\textbf{Supplemental Material}}
\end{center}

\section{Evaluation of the speed limits}
Here, we explain the derivation of Eq.~\eqref{eq.bound.eval}. 
Taking the difference between $v_{\rm info}$ and $v_R^{\lambda}+v_{\rm info}^{\mathsf{W}}$, we get $v_{\rm info}-\qty(v_R^{\lambda}+v_{\rm info}^{\mathsf{W}})=v_{\rm info}-v_{\rm info}^{\mathsf{W}}-\cos\theta_R^{\lambda}v_{\rm info}^{\lambda}$. Thus, the following equations hold. 
\begin{align}
    \cos\theta_R^{\lambda}\leq  \dfrac{v_{\rm info}-v_{\rm info}^{\mathsf{W}}}{v_{\rm info}^{\lambda}} &\implies (v_R\leq\;)\;v_R^{\lambda}+v_{\rm info}^{\mathsf{W}}\leq v_{\rm info},\notag\\
    \cos\theta_R^{\lambda}\geq \dfrac{v_{\rm info}-v_{\rm info}^{\mathsf{W}}}{v_{\rm info}^{\lambda}} &\implies (v_R\leq\;)\;v_{\rm info}\leq v_R^{\lambda}+v_{\rm info}^{\mathsf{W}}.
\end{align}
The same calculations can be applied to the lower limits: $-v_{\rm info}$ and $v_R^{\lambda}-v_{\rm info}^{\mathsf{W}}$. The difference between the two is, $\qty(v_R^{\lambda}-v_{\rm info}^{\mathsf{W}})-(-v_{\rm info})=v_{\rm info}+\cos\theta_R^{\lambda}v_{\rm info}^{\lambda}-v_{\rm info}^{\mathsf{W}}$. Therefore,
\begin{align}
    \cos\theta_R^{\lambda}\geq -\dfrac{v_{\rm info}-v_{\rm info}^{\mathsf{W}}}{v_{\rm info}^{\lambda}}&\implies
    -v_{\rm info}\leq v_R^{\lambda}-v_{\rm info}^{\mathsf{W}}\;(\;\leq v_R),\notag\\
    \cos\theta_R^{\lambda}\leq -\dfrac{v_{\rm info}-v_{\rm info}^{\mathsf{W}}}{v_{\rm info}^{\lambda}}&\implies
    v_R^{\lambda}-v_{\rm info}^{\mathsf{W}}\leq -v_{\rm info}\;(\;\leq v_R).
\end{align}
Taking the $\arccos$ on both sides of these equations and using the relation $\arccos(-x) = \pi-\arccos(x)$, we obtain Eq.~\eqref{eq.bound.eval} in the main text.

\section{Detail of the mutation bound}
In this section, we describe the results of the numerical calculation for the following speed limit, which we call the mutation bound:
\begin{align}
     v_R^{\mathsf{W}}-v_{\rm info}^{\lambda}\leq v_R \leq v_R^{\mathsf{W}} + v_{\rm info}^{\lambda}.
\end{align}
This speed limit is expected to give a good evaluation in mutation-dominant situations, whereas the selection bound gives a good evaluation in selection-dominant situations. We below proceed with the discussion in parallel with that in the main text. We define the angle $\theta_R^{\mathsf{W}}$ between observable $R$ and the stochastic entropy change rate in the system due to mutation $-\dot\sigma^{\mathsf{W}}$ as
\begin{align}
    \theta_R^{\mathsf{W}} = \arccos \left(\frac{\langle\Delta R,-\dot\sigma^{\mathsf{W}}\rangle}{\Vert \Delta R\Vert \Vert-\dot{\sigma}^{\mathsf{W}}\Vert} \right) =  \arccos \left(\frac{v_R^{\mathsf{W}}}{v_{\rm info}^{\mathsf{W}}} \right).
\end{align}
Using this angle $\theta_R^{\mathsf{W}}$ and another angle $\theta^\dag$ defined as 
\begin{align}
    \theta^{\dag} \coloneqq \arccos\qty(\dfrac{v_{\rm info}-v_{\rm info}^{\lambda}}{v_{\rm info}^{\mathsf{W}}}), 
\end{align}
the relations among the upper bound $v_{\rm info}^{\lambda}+v_R^{\mathsf{W}}$ and the lower bound $-v_{\rm info}^{\lambda}+v_R^{\mathsf{W}}$ by the mutation bound and $v_{\rm info}$ are expressed as
\begin{align}
    \begin{cases}
        0\leq \theta_R^{\mathsf{W}}\leq \theta^\dag &\implies v_R\leq v_{\rm info}\leq  v_R^{\mathsf{W}}+v_{\rm info}^{\lambda}\\
        \theta^\dag\leq \theta_R^{\lambda}\leq \pi &\implies v_R\leq v_R^{\mathsf{W}}+v_{\rm info}^{\lambda} \leq v_{\rm info}  \\
        0\leq \theta_R^{\lambda}\leq \pi-\theta^\dag &\implies -v_{\rm info} \leq v_R^{\mathsf{W}}-v_{\rm info}^{\lambda}\leq v_R\\
        \pi -\theta^\dag\leq \theta_R^{\lambda}\leq \pi &\implies v_R^{\mathsf{W}}-v_{\rm info}^{\lambda}\leq -v_{\rm info}\leq v_R 
    \end{cases}.
\end{align}
In contrast to $\theta^\ast$, $\theta^\dag$ becomes small when the contribution of mutation to the evolution of the probability distribution is large. It indicates that the range of $\theta_R^{\mathsf{W}}$ where the mutation bound is tighter, $[\theta^\dag,\pi-\theta^\dag]$, gets wider. To sum up, the selection bound gets tighter when the contribution of selection is large, while the mutation bound gets tighter when the contribution of mutation is large.

\begin{figure}[ht]
  \begin{minipage}[t]{0.45\linewidth}
    \centering
    \includegraphics[keepaspectratio, scale=0.27]{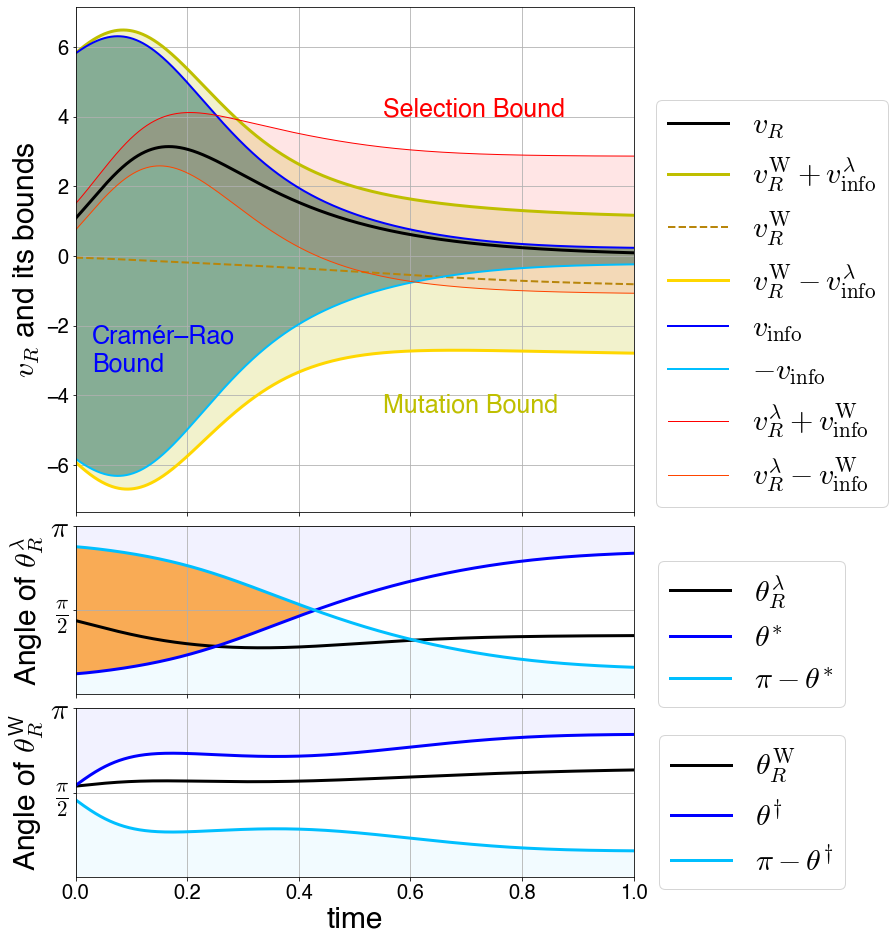}
    \caption{Numerical calculation of a situation where the selection bound can be tighter for $v_R$ than Cram\'{e}r--Rao bound and the mutation bound cannot. The horizontal axis is time $t$ and each value is plotted in the upper part of the graph. The lower part shows the angles. The number of species are $n=10$, the observable $R$, growth rate $\lambda$, and the off-diagonal components of the transition matrix $\mathsf{W}$ are taken as uniform random numbers in $[-10,10],[-10,10], \mathrm{and } [0.0.1]$, respectively. The initial distribution is also generated as uniform random numbers.}
    \label{fig:Mbound_1}
  \end{minipage}
  \hfill
  \begin{minipage}[t]{0.45\linewidth}
    \centering
    \includegraphics[keepaspectratio, scale=0.27]{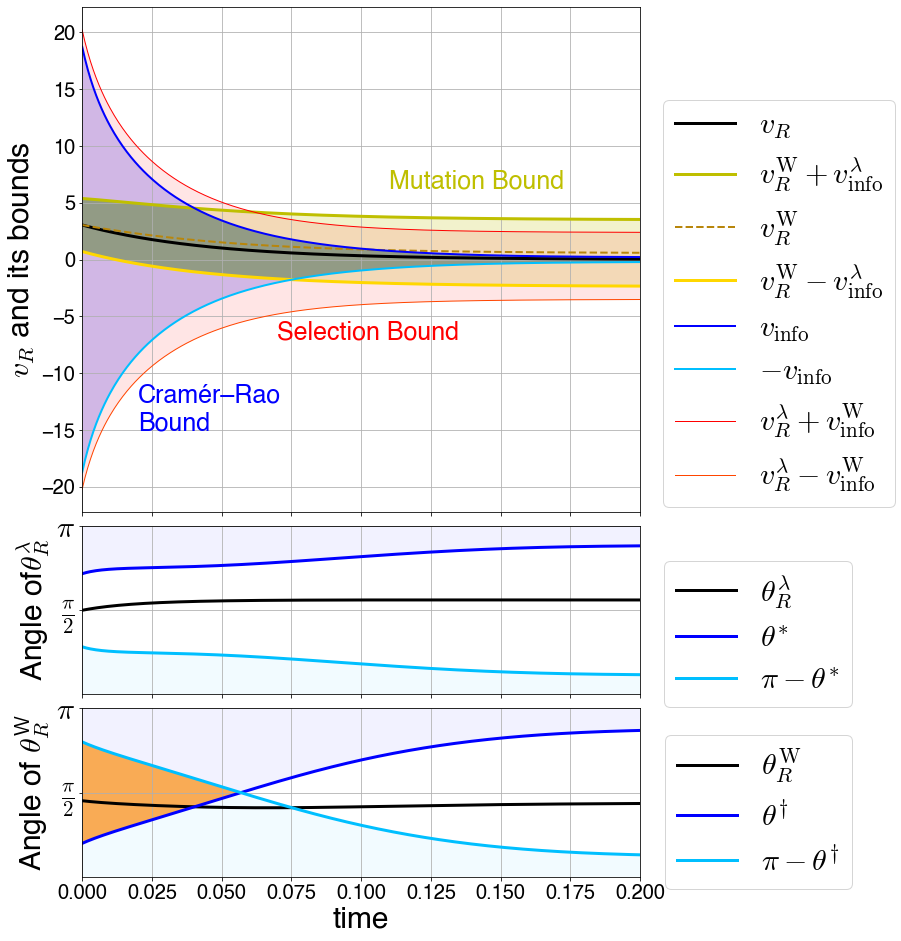}
    \caption{Numerical calculation of a situation where the mutation bound can be tighter for $v_R$ than Cram\'{e}r--Rao bound and the selection bound cannot. The horizontal axis is time $t$ and each value is plotted in the upper part of the graph. The lower part shows the angles. Compared to Fig.~\ref{fig:Mbound_1}, the parameters are different. The observable $R$, growth rate $\lambda$, transition rate $\mathsf{W}$ is generated as uniform random numbers in $[-10,10], [-5,5], \mathrm{and } [0,5]$, respectively. Note that the ratio of the growth rate to the mutation rate is $1$, not $100$ as in Fig.~\ref{fig.NS.graph} in the main text and Fig.~\ref{fig:Mbound_1}.}
    \label{fig:Mbound_2}
  \end{minipage}
\end{figure}

With the parameters used to demonstrate the selection bound in the main text, the mutation bound gives a loose bound. As shown in Figs.~\ref{fig:Mbound_1}~and~\ref{fig:Mbound_2}, the mutation bound is loose in situations where the selection bound gives a tight evaluation, and conversely, the selection bound is loose in situations where the mutation bound is tight. 

Thus, the selection bound and the mutation bound provide good bounds to evaluate the change speed of the observables in different situations. 

\section{How the strength of selection and mutation affect the evaluation of the speed limits}
In order to evaluate the tightness of the selection bound and the mutation bound quantitatively, we discuss the dependence of the ``tendency'' to give better limits on the ratio of $\lambda_{\rm max}$ to $\mathsf{W}_{\rm max}$. 
This tendency for the selection bound to be tighter than the Cram\'{e}r--Rao bound can be measured using the value of $\theta^\ast$. The selection bound gets tighter when the angle $\theta_R^\lambda$ is in the range $[\theta^\ast,\pi-\theta^\ast]$. Therefore, if we define $\mathcal{P}^\ast\coloneqq \max\{(\pi-2\theta^\ast)/\pi,0\}$, this $\mathcal{P}^\ast$ quantifies the tendency of the selection bound to give a better evaluation. 
The selection bound tends to be tighter when $\mathcal{P}^\ast$ is close to $1$ and looser when $\mathcal{P}^\ast$ is close to $0$. 
Note that $\theta^\ast$ is not always under $\pi/2$, thus, $\pi-2\theta^\ast$ could be negative. In the same way, the tendency of the mutation bound to give a better evaluation can be measured by the value defined as $\mathcal{P}^\dag\coloneqq \max\{(\pi-2\theta^\dag)/\pi,0\}$. It is noteworthy that $\mathcal{P}^\ast$ and $\mathcal{P}^\dag$ do not depend on the observable $R$. 

\begin{figure}[t]
    \centering
    \includegraphics[scale=0.45]{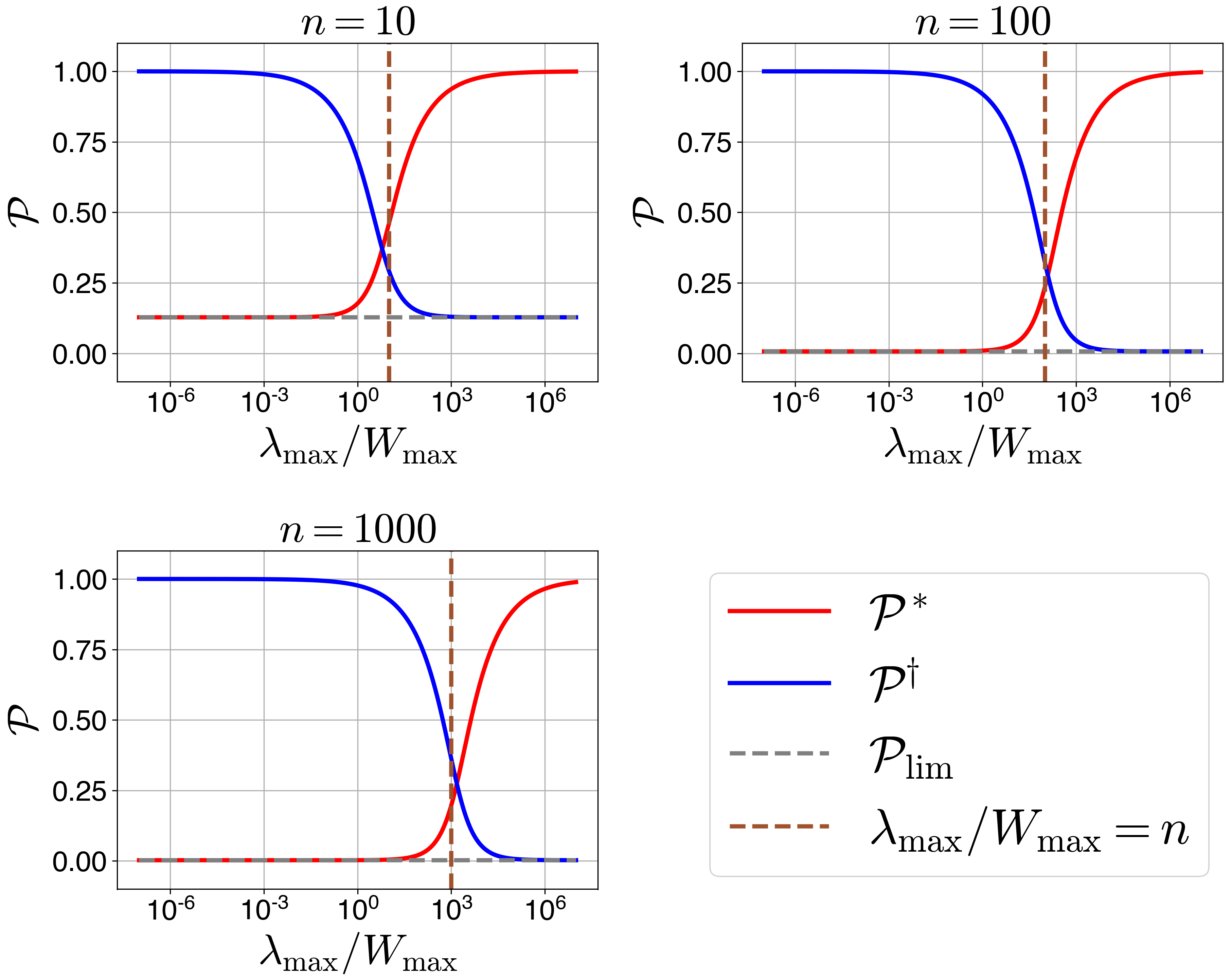}
    \caption{Numerical calculation for $\mathcal{P}^\ast$ and $\mathcal{P}^\dag$, that quantify the tendency of the selection/mutation bound to get tighter than the Cram\'{e}r--Rao bound. $\lambda$ and $\mathsf{W}$ are generated by uniform random numbers in $[-\lambda_{\rm max},\lambda_{\rm max}]$ and $[0,\mathsf{W}_{\rm max}]$, respectively. The graphs show how $\mathcal{P}^\ast$ and $\mathcal{P}^\dag$ depend on the ratio of $\lambda_{\rm max}$ to $\mathsf{W}_{\rm max}$. $\mathcal{P}_{\rm lim}$ is the limit of $\mathcal{P}^\ast$ as $\lambda_{\rm max}/\mathsf{W}_{\rm max}\to 0$ or  that of $\mathcal{P}^\dag$ as $\lambda_{\rm max}/\mathsf{W}_{\rm max}\to \infty$. These two values converge to an identical value, which does not necessarily go to $0$. Each graph is the result for a different number of species $n=10,100,1000$. }
    \label{fig:SM_LWrate}
\end{figure}

We demonstrate the dependence of $\mathcal{P}^\ast$ and $\mathcal{P}^\dag$ on the ratio of $\lambda_{\rm max}$ to $\mathsf{W}_{\rm max}$ (see Fig.~\ref{fig:SM_LWrate}). In the numerical calculation, $\lambda$ and $\mathsf{W}$ is generated by uniform random values in the range $[-\lambda_{\rm max},\lambda_{\rm max}]$, and $[0,\mathsf{W}_{\rm max}]$, respectively. 
Note here that both $\mathcal{P}^\ast$ and $\mathcal{P}^\dag$ depend on time, so we use their values in the initial state of the dynamics. 
Fig.~\ref{fig:SM_LWrate} shows that the tendency $\mathcal{P}^\ast$ of the selection bound to get tighter than the Cram\'{e}r--Rao bound increases when $\lambda_{\rm max}$ is larger than $\mathsf{W}_{\rm max}$. This calculation confirms the statement in the main text that the selection bound is effective when natural selection is dominant.

Another finding is that the point where the two curves of $\mathcal{P}^\ast$ and $\mathcal{P}^\dag$ intersect is governed by the number of species $n$. 
This point represents the ratio $\lambda_{\rm max} / \mathsf{W}_{\rm max}$ with $v_{\rm info}^\lambda= v_{\rm info}^{\mathsf{W}}$. Moreover, the tendency for the selection/mutation bound to give better evaluation swaps at this point. Fig.~\ref{fig:SM_LWrate} shows that such a ratio is not $1$, but close to the number of species $n$. It implies that the contribution of selection and that of mutation compete when $\lambda_{\rm max}$ is about $n$ times larger than $\mathsf{W}_{\rm max}$. This fact can be understood by considering a situation where the parameters are $\lambda_i=1,\;W_{ij}=1(i\neq j)$. The contribution to the growth rate of trait $i$ of the selection is $1$, whereas the contribution of the mutation is $n-1$. Therefore, if the traits grow at an identical rate, the contribution of selection becomes larger in small communities.

\section{\red{How to verify the selection bound and the mutation bound from experimental data}}

%The ``observables'' in our Letter may not be precisely measured in actual experiments because cell proliferation during a finite period of measurement time introduces systematic errors in the experimental data at population level~\cite{hashimoto2016noise}. However, by experimentally obtaining single-cell lineage tree data, one can eliminate such effects and can precisely estimate the instantaneous values of quantities such as $p_i(t)$ and fitness variance~\cite{nozoe2017inferring}. 

%One can measure when and into what each cell mutates or divides, for example, by analyzing time-lapse images of growing bacteria~\cite{nozoe2017inferring}. The traits are either genetic or phenotypic (e.g., cell sizes or shapes and intracellular concentration of a particular protein if probed with an appropriate fluorescence reporter).
%In this way, the single-cell lineage tree structure can be reconstructed, showing genealogical relationships among the individual cells that originated from ancestor cell. 

% only the information available from the cell lineages obtained from the experimental systems capable of measuring the number of divisions of an individual~\cite{wakamoto2005single, hashimoto2016noise}. 

\begin{figure}[t]
    \centering
    \includegraphics[scale=0.5]{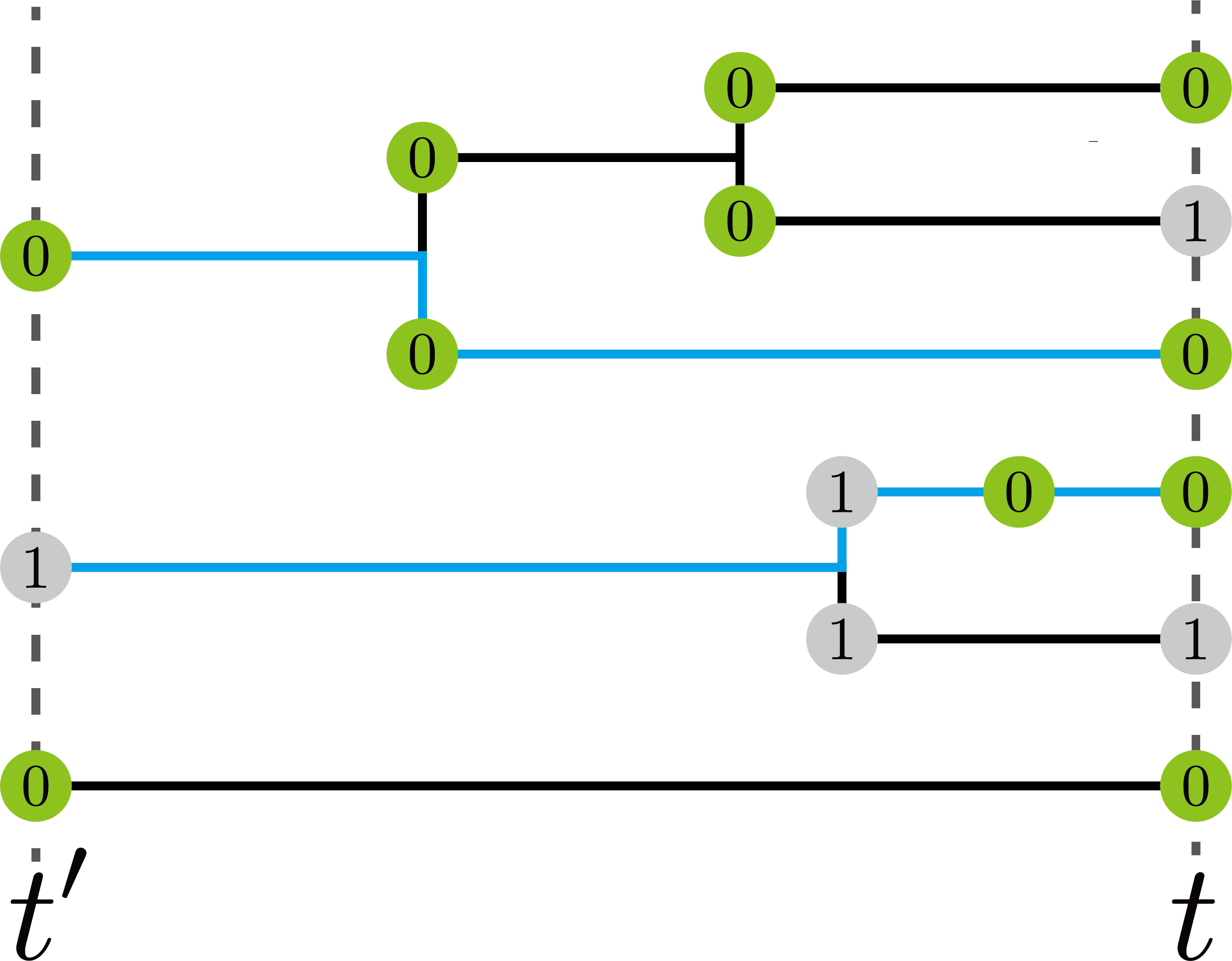}
    \caption{\red{
    Example of single-cell lineage trees with two traits, $0$~(green) and $1$~(gray). The distributions $p(t)$ and $p^{\rm ch}(t)$ are available only from the cell lineage by counting $N_{i}(K,t;t')$. The blue lines show the trajectories of individuals that experienced $1$ divisions between time $t'$ and $t$ in the subpopulation with the trait $0$ at time $t$. Thus $N_0(1,t;t')$ is $2$ in this cell lineage.}
    }
    \label{celllineage}
\end{figure}

\red{In this section, we describe how one can quantitatively test our speed limit using single-cell genealogical data in the form of population lineage trees which include data of cell divisions and mutations or phenotypic switching (see Fig.~\ref{celllineage} as an example). We can compute the number $N_i(K,t;t')$ of individuals with $i$-th trait at time $t$ that have experienced $K$ divisions since time $t'$ from single-cell genealogical data. $N_i(K,t;t')$ enable us to compute the instantaneous values of all the quantities in our speed limit, i.e., $p_i(t)$, ${\dot\sigma}_i^{\lambda}$, ${\dot\sigma}_i^{\mathsf{W}}$, and ${\dot\sigma}_i^{\lambda}$ as well as $v_{R}^{\lambda}, v_{R}^{\mathsf{W}}, v_{\rm info}^{\lambda}$ and $v_{\rm info}^{\mathsf{W}}$ for an arbitrary observable $R$. The details are as follows.}

\red{Firstly, from $N_i(K,t;t')$, we define chronological(forward) distribution $p^{\rm ch}$ as
\begin{align}\label{fw_def}
 p_{i}^{\rm ch}(t;t')\coloneqq\sum_{K=0}^{\infty}\frac{N_i(K,t;t')}{N_{\rm tot}(t')2^K}.
\end{align}
From these values, we can obtain the instantaneous values of $N_i(t)$ and $p_i(t)$ for each subpopulation: 
the sum of $N_i(K,t;t')$ with respect to $K$ equals $N_i(t)$,
\begin{align}\label{sum_property}
    \sum_{K=0}^{\infty}N_i(K,t;t')=N_i(t), 
\end{align}
and $p_{i}^{\rm ch}(t;t')$ is equal to $p_i(t)$ when $t'=t$,
\begin{align}\label{same_time_condition}
    p_{i}^{\rm ch}(t;t)=p_i(t),
\end{align}
because only $K=0$ is allowed as the number of division between $t$ and $t$. % Thus, $p_i(t)$ and $p_{i}^{\rm ch}(t;t')$ are available. 
Note that the chronological distribution defined above has been studied not only as a theoretical object~\cite{baake2007mutation,kobayashi2015fluctuation} but also utilized to analyze experimental data as in Refs.~\cite{nozoe2017inferring,genthon2020fluctuation}. $t'$ is usually set to the initial time and not explicitly written in these previous studies.}

\red{Up to here, we can compute speeds $v_R^\lambda$, $v_R^\mathsf{W}$, $v_\mathrm{info}^\lambda$ and $v_\mathrm{info}^\mathsf{W}$ for any observable $R$ by using $p_i(t)$. 
On the other hand, since the remaining quantities, $\dot{\sigma}_i$, $\dot{\sigma}_i^\lambda$, and $\dot{\sigma}_i^\mathsf{W}$, depend on the other parameters, $\lambda$ and $\mathsf{W}$, we need more calculations. 
Surprisingly, single-cell genealogical data enable us to compute them without estimating the parameters, as we show below.}

\red{As a preparation, let us consider the differential equations that $N_i(K,t;t')$ and $p^\mathrm{ch}$ satisfy. 
Let an individual with $i$th trait divide into $2$ individuals at rate $r_i\geq0$. 
Then, by division, $N_i(K,t;t')$ increases by $r_iN_i(K-1,t;t')dt$ in an infinitesimal duration $dt$, while decreasing by $r_iN_i(K,t;t')dt$. Note that only one division can occur in an infinitesimal duration $dt$.
By taking into account the term due to mutation $\sum_{j=1}^{n}W_{ij}N_j(K,t;t')$, we get the equation
\begin{align}\label{time_development_of_NK}
    \frac{d}{dt}N_i(K,t;t')=2r_iN_i(K-1,t;t')-r_iN_i(K,t;t')+\sum_{j=1}^{n}W_{ij}N_j(K,t;t'),
\end{align}
where we define $N_i(K,t;t')=0$ for $K<0$. 
% Assuming Eq.~(\ref{time_development_of_NK}) is reasonable because we can obtain the model equivalent to the one used in the paper. In fact, 
By taking the sum for $K$, we obtain
\begin{align}\label{time_development_of_bw}
    \frac{d}{dt}N_i(t)=r_iN_i(t)+\sum_{j=1}^{n}W_{ij}N_j(t). 
\end{align}
If we write $r_i$ as $\lambda_i$, it is identical to our model in the main text. Note that $r_i$ is all non-negative, while $\lambda_i$ can be negative values in general, which reflects the experimental setup that does not account for individual mortality. 
On the other hand, combining Eq.~(\ref{time_development_of_NK}) with Eq.~(\ref{fw_def}), we find
\begin{align}
    \frac{d}{dt}p^{\rm ch}_i(t;t')&=\sum_{K=0}^{\infty}\frac{1}{N_{\rm tot}(t')2^K}\qty[2r_iN_i(K-1,t;t')-r_iN_i(K,t;t')+\sum_{j=1}^{n}W_{ij}N_j(K,t;t')]\notag\\
    &=r_i\sum_{K=0}^{\infty}\frac{N_i(K-k_i,t;t')}{N_{\rm tot}(t')2^{K-1}}-
    r_i\sum_{K=0}^{\infty}\frac{N_i(K,t;t')}{N_{\rm tot}(t')2^K}+\sum_{j=1}^{n}W_{ij}\sum_{K=0}^{\infty}\frac{N_j(K,t;t')}{N_{\rm tot}(t')2^K}\notag\\
    &=\sum_{j=1}^{n}W_{ij}p^{\rm ch}_j(t;t').\label{time_development_of_fw}
\end{align}
}
\red{Now we can present the way to compute ${\dot\sigma}_i^{\mathsf{W}}$ and ${\dot\sigma}_i^{\lambda}$, using only the distributions $p_i(t)$ and $p_{i}^{\rm ch}(t;t')$, which are available from single-cell lineage tree data. 
Discretizing Eq.~(\ref{time_development_of_fw}) and dividing the obtained equation by $p_i(t)$, we find the relation
\begin{align}\label{discrete_dt_fw}
    \frac{p_i^{\rm ch}(t+dt;t')-p_i^{\rm ch}(t;t')}{p_i(t)dt}=\sum_{j=1}^{n}W_{ij}\frac{p^{\rm ch}_j(t;t')}{p_i(t)}.
\end{align}
Then, substituting $t$ for $t'$ and using Eq.~(\ref{same_time_condition}), we see that we can calculate ${\dot\sigma}_i^{\mathsf{W}}$ as 
\begin{align}\label{W_ent}
    \frac{p_i^{\rm ch}(t+dt;t)-p_i(t)}{p_i(t)dt}=\sum_{j=1}^{n}W_{ij}\frac{p_j(t)}{p_i(t)}=-{\dot\sigma}_i^{\mathsf{W}}.
\end{align}
On the other hand, if we consider the discretization of Eq.~(\ref{time_development_of_bw}), as we did for Eq.~(\ref{time_development_of_fw}), we obtain ${\dot\sigma}_i$ as 
\begin{align}\label{ent}
    \frac{p_i(t+dt)-p_i(t)}{p_i(t)dt}=\Delta\lambda_i+\sum_{j=1}^{n}W_{ij}\frac{p_j(t)}{p_i(t)}=-{\dot\sigma}_i.
\end{align}
Then ${\dot\sigma}_i^{\lambda}$ is given by ${\dot\sigma}_i^{\lambda}={\dot\sigma}_i-{\dot\sigma}_i^{\mathsf{W}}$. As a result, it is finally shown that we can compute all the relevant quantities we discuss in the main text from single-cell genealogical data.}

\end{document}